\documentclass[pra,twocolumn,showpacs,superscriptaddress,aps]{revtex4-1}
\usepackage{amsmath}
\usepackage{amsfonts}
\usepackage{amssymb}
\usepackage{physymb}
\usepackage{graphicx}
\usepackage{epsfig}

\usepackage{dcolumn}
\usepackage{bm}
\usepackage{fancybox}
\usepackage{multirow}
\usepackage{float}
\usepackage{hyperref}
\usepackage{epstopdf}

\usepackage[utf8]{inputenc}

\newcommand{\Punkte}{0}

{\noindent {\bf Exercise {#1}.} #2 \vspace{0.2cm} \\
\renewcommand{\Punkte}{#3}}%
{\mbox{\hspace{3ex}} \hfill {\bf \Punkte~\mbox{Points}}\bigskip }

\newenvironment{Exercise*}[2]%
{\noindent {\bf Exercise {#1}*.} #2 \vspace{0.2cm} \\
renewcommand{\Punkte}{#2}}%
{\mbox{\hspace{2ex}} \hfill {\bf \Punkte~\mbox{Points}}\bigskip }

{\noindent {\bf Exercise {#1}.}\vspace{0.2cm} #2}%
{}

\newcounter{enum1}

\newcounter{enuma}

\begin{document}

\title{{\bf When Does Linear Stability Not Exclude Nonlinear Instability~?}}
\date{\today}

\author{P. G. Kevrekidis \thanks{%
Email: kevrekid@math.umass.edu}}
\affiliation{Department of Mathematics and Statistics, University of Massachusetts,
Amherst, MA 01003-4515, USA}

\affiliation{Center for Nonlinear Studies and Theoretical Division, Los Alamos
National Laboratory, Los Alamos, NM 87544}

\author{D. E. Pelinovsky}
\affiliation{Department of Mathematics, McMaster University, Hamilton, Ontario, Canada, L8S 4K1}

\author{A. Saxena}
\affiliation{Center for Nonlinear Studies and Theoretical Division, Los Alamos
National Laboratory, Los Alamos, NM 87544}

\begin{abstract}
We describe a mechanism that results in the
nonlinear instability of stationary states {\it even in the case where the
stationary states are linearly stable}. This instability is due to the nonlinearity-induced coupling
of the linearization's internal modes of
negative energy with the wave continuum.
In a broad class of nonlinear Schr{\"o}dinger (NLS)
equations considered, the presence of such internal modes {\it guarantees}
the nonlinear instability of the stationary states in the evolution dynamics.
To corroborate this idea, we explore three prototypical case examples:
(a) an anti-symmetric soliton in a double-well potential,
(b) a twisted localized mode in a one-dimensional lattice with cubic nonlinearity,  and
(c) a discrete vortex in a two-dimensional saturable lattice.
In all cases, we observe a weak nonlinear instability, despite the
linear stability of the respective states.
\end{abstract}

\maketitle

{\it Introduction.} Among dispersive nonlinear partial
differential equations, the nonlinear Schr\"odinger (NLS) model \cite{ablowitz1,sulem}
stands out as a prototypical system that has proved to be
essential in modeling and understanding features of numerous areas in nonlinear physics.
The relevant fields of application vary from optics and the propagation of the electric
field envelope in optical fibers \cite{hasegawa,kivshar}, to
the self-focusing and collapse of Langmuir waves in plasma physics
\cite{zakh1,zakh2}. It is also encountered in the modeling of
deep water and freak/rogue waves in the
ocean \cite{benjamin,onofrio},
as well as in atomic physics and the dynamics of superfluids and atomic
Bose-Einstein condensates~\cite{becbook1,becbook2,rcg:BEC_BOOK}.

One of the most customary ways to approach the experimental observations
of nonlinear dispersive waves in these different physical systems is to explore
the standing wave solutions that NLS models may possess and to understand
their spectral and dynamical stability characteristics~\cite{toddkeith}.
This is accomplished not only in homogeneous continuous media, but also in
inhomogeneous and discrete ones, not only in one but
also in higher dimensions~\cite{jyang,pelinov,dnlsbook}. Then, the
conventional wisdom suggests that should the solution in the NLS model
be found to be linearly (spectrally) stable, then
it should be expected to be dynamically stable as well and hence
a suitable candidate for observations in physical experiments.
By linear stability here, we imply the absence of eigenvalues with
nonzero real parts, as well as the absence of multiple and embedded
imaginary eigenvalues with the exception of the zero eigenvalue generated
by the symmetries of the NLS models.

In the present work, we explore an important, as well as {\it generic}
nonlinear mechanism of
instability of standing wave solutions, which are linearly stable.
The instability is induced by the linearization's
internal modes of {\it negative energy} or
{\it negative Krein signature}~\cite{kapkev} that correspond
to simple imaginary eigenvalues and represent negative ``directions"
of the NLS energy at the standing wave solutions. While perfectly innocuous
in the linear setting, these internal modes of negative energy can be in resonance with the wave spectrum
or other internal modes due to nonlinearity, in which case they lead to the
nonlinear instability of the standing wave solutions.

For a ground state, small amplitude excitations of the standing wave always
increase the NLS energy so that the standing wave is an {\it energetically} stable
minimum of the system, which is also {\it dynamically} stable. Internal modes
for such ground states may only have positive energy or positive Krein signature \cite{kiv1,kiv2}.
However, for many excited states, small amplitude excitations may decrease
the NLS energy and still do not result in the appearance of linear instability.
This has led to the widespread belief
that energetic
instability does not generically imply dynamical instability;
instead ``the energetic instability can only destabilize the
system in the presence of dissipative terms which drive it towards
configurations of lower energy'' (p. 58 in~\cite{becbook1}).
Our aim herein is to challenge this conventional
wisdom and to establish through a diverse array of
case examples that, in fact, generically, {\it excited states
bearing an energetic instability, will also manifest
a dynamical one}.

The initial mathematical formulation of the nonlinear mechanism of
instability of excited states was obtained by Cuccagna~\cite{scipio,scipio1},
but these results have not been confirmed  in physics literature by
numerical or experimental evidence. In the present work, we give
convincing numerical evidence of the nonlinear instability due to the
internal modes of negative energy based on three case examples involving
continuous and discrete NLS in one and two dimensions. The discovery of this
nonlinear mechanism may broadly impact researchers in nonlinear physics
enabling them to identify and to explain
weak (nonlinear) instabilities of the standing waves observed in
numerous experimental setups in atomic, optical, fluid or plasma systems
related to this general framework.

{\it Theoretical Formulation.} The cubic NLS model in a generalized form
reads:
\begin{eqnarray}
i \partial_t u = - \nabla^2 u + V(x) u + g |u|^2 u,
\label{nls}
\end{eqnarray}
where $u$ is a complex field, $V$ characterizes the external potential
with a fast decay to zero at infinity,
and $g$ is a coefficient that characterizes
the self-focusing ($g<0$) or self-defocusing ($g>0$)
nature of the nonlinearity. The standing wave solution
takes the form $u(x,t) = e^{-i \omega t} \phi(x)$,
where $\omega$ and $\phi$ are real.

The prototypical example of the nonlinear instability in the NLS
model (\ref{nls}) occurs when the Schr\"{o}dinger operator $-\nabla^2 + V$
admits two simple negative eigenvalues (energy levels) $E_0 < E_1$
such that
\begin{equation}
\label{eig}
\frac{1}{2} |E_1| < |E_1 - E_0| < |E_1|.
\end{equation}
The ground state bifurcates for $\omega$ near $E_0$, whereas the excited
state bifurcates for $\omega$ near $E_1$. In particular, the excited state
can be represented by $\phi(x) = \epsilon u_1(x) + \mathcal{O}(\epsilon^3)$, where
$u_1$ is the $L^2$-normalized eigenfunction of $-\nabla^2 + V$ for eigenvalue $E_1$,
and $\epsilon$ is found from $\omega$ by
$$
\omega = E_1 + g \left( \int_{\mathbb{R}} |u_1|^4 dx \right) \epsilon^2.
$$

When considering the stability of the standing wave, we utilize the linearization, e.g.
by means of
\begin{eqnarray}
u(x,t)= e^{-i \omega t} \left[\phi(x) + \delta
\left(a(x) e^{\lambda t} + \bar{b}(x) e^{\bar{\lambda} t} \right) \right],
\label{eqn1}
\end{eqnarray}
with small parameter $\delta$ (independently of $\epsilon$),
and obtain the spectral problem
\begin{eqnarray}
H \psi = i \lambda \sigma_3 \psi,
\label{spectrum}
\end{eqnarray}
where $\psi = (a,b)^T$, $\sigma_3 = {\rm diag}(1,-1)$, and $H$ is given by
$$
H = \left[ \begin{array}{cc} - \Delta + V - \omega + 2 g \phi^2 & g \phi^2 \\
g \phi^2 & - \Delta + V - \omega + 2 g \phi^2 \end{array} \right].
$$
If $\epsilon$ is small, it is easy to confirm that the spectral problem (\ref{spectrum})
has a double zero eigenvalue due to the gauge symmetry of the NLS model,
the phonon band for $\lambda \in i (-\infty,-|\omega|]$ and $\lambda \in i [|\omega|,\infty)$,
and a pair of internal modes at $\lambda = \pm i \Omega$ with $\Omega = E_1 - E_0 + \mathcal{O}(\epsilon^2) > 0$.
With the condition (\ref{eig}), we note that $\Omega < |\omega|$ but $2 \Omega > |\omega|$,
hence the internal mode eigenfrequency is isolated from the phonon band but the second harmonic
is embedded into the phonon band. The second harmonic can be generated by the
nonlinear terms beyond the linear approximation (\ref{eqn1}). Also
note that the mode energy is defined by
\begin{eqnarray}
K = \langle H \psi_{\Omega}, \psi_{\Omega} \rangle =  - \Omega \int_{\mathbb{R}} (|a_{\Omega}|^2 - |b_{\Omega}|^2) dx,
\label{eqn2}
\end{eqnarray}
where $\psi_{\Omega} = (a_{\Omega},b_{\Omega})^T$ is the eigenvector for the eigenvalue $\lambda = i \Omega$.
Since  $\Omega > 0$ and $\psi_{\Omega} = (u_0,0)^T + \mathcal{O}(\epsilon^2)$, where $u_0$ is the $L^2$-normalized
eigenfunction of $-\nabla^2 + V$ for eigenvalue $\lambda_0$,
then it follows that the internal mode has negative energy, that is, $K < 0$.

Now we will explain why the internal mode of negative energy, when coupled with the phonon band due to the second harmonic, leads to the nonlinear instability of the standing wave.
We shall consider the expansion in amplitudes of the internal mode
\begin{eqnarray*}
u(x,t) = e^{-i \omega t} \left[\phi(x) + \delta u_1(x,t) + \delta^2 u_2(x,t) + \mathcal{O}(\delta^3) \right],
\end{eqnarray*}
with 
\begin{eqnarray*}
u_1(x,t) = c(\tau) a_{\Omega}(x) e^{i \Omega t} +
\bar{c}(\tau) \bar{b}_{\Omega}(x) e^{- i \Omega t},
\end{eqnarray*}
where $c(\tau)$ is the complex amplitude of the internal mode evolving in 
slow time $\tau = \delta^2 t$. Note that because $H + \Omega \sigma_3$ is a self-adjoint operator, 
we can choose the internal mode $(a_{\Omega},b_{\Omega})^T$ to be real. 
Using the expansion above, similar to what was done in the case of internal modes
of positive energy \cite{kiv2}, we obtain the explicit representation of
the second-order correction term
\begin{eqnarray*}
u_2(x,t) = c^2 a_2(x) e^{2 i \Omega t} +
|c|^2 a_0(x) + \bar{c}^2 \bar{b}_2(x) e^{- 2i \Omega t},
\end{eqnarray*}
where $\psi_2 = (a_2,b_2)^T$ and $\psi_0 = (a_0,a_0)^T$ are obtained from suitable
solutions of the inhomogeneous problems
\begin{eqnarray}
(H + 2 \Omega \sigma_3) \psi_2 = -g \phi \left[ \begin{array}{c} (a_{\Omega}+2b_{\Omega}) a_{\Omega} \\
(2a_{\Omega}+b_{\Omega}) b_{\Omega} \end{array} \right]
\label{inhom-eq0}
\end{eqnarray}
and
\begin{eqnarray}
H  \psi_0 = -2 g \phi (a_{\Omega}^2 +  a_{\Omega} b_{\Omega} + b_{\Omega}^2)
\left[ \begin{array}{c} 1 \\ 1 \end{array} \right].
\label{inhom-eq}
\end{eqnarray}
Because  $2 \Omega > |\omega|$, the correction term $\psi_2$ is bounded but not decaying at infinity.
We apply the Sommerfeld radiation condition
\begin{equation}
\label{Sommerfeld}
\psi_2(x) \to R_{\Omega} e_2 e^{\mp i k_{\Omega} x} \quad \mbox{\rm as} \quad x \to \pm \infty,
\end{equation}
where $e_2 = (0,1)$, $k_{\Omega} = \sqrt{2 \Omega - |\omega|}$, and
the radiation tail amplitude $R_{\Omega}$ is a uniquely determined complex coefficient.
Because of the Sommerfeld radiation condition (\ref{Sommerfeld}), 
the second harmonic $(a_2,b_2)$ is given by complex functions. 

Proceeding to the third-order correction term, as in \cite{kiv2}, we obtain
the evolution equation for the amplitude $c(\tau)$ in slow time $\tau = \delta^2 t$,
\begin{equation}
\label{normal-form}
i K \frac{dc}{d\tau} + \Omega \beta |c|^2 c = 0,
\end{equation}
where $K$ is given by (\ref{eqn2}) and $\beta$ is found from the projections as
\begin{eqnarray*}
\beta & = &  2 g \int_{\mathbb{R}} \phi \left[ a_2 a_{\Omega} (a_{\Omega}+2b_{\Omega}) + b_2 b_{\Omega}(2a_{\Omega}+b_{\Omega}) \right] dx \\
& \phantom{t} & + 4 g \int_{\mathbb{R}} a_0 (a_{\Omega}^2 + a_{\Omega} b_{\Omega} + b_{\Omega}^2) dx \\
& \phantom{t} & +  g \int_{\mathbb{R}} \left( a_{\Omega}^4 + 4 a_{\Omega}^2 b_{\Omega}^2 + b_{\Omega}^4 \right) dx.
\end{eqnarray*}
Note that the coefficients in front of $(a_2,b_2)$ correspond to the source term
of the inhomogeneous equations (\ref{inhom-eq0}). The parameter $\beta$ is complex because
the correction terms $(a_2,b_2)$ are complex. Using Eqs. (\ref{inhom-eq0}) and
(\ref{Sommerfeld}), we obtain
\begin{eqnarray}
\label{Fermi-rule}
2 i {\rm Im}(\beta) &=& 2 \left( \bar{a}_2' a_2 + \bar{b}_2' b_2 - \bar{a}_2 a_2' - \bar{b}_2 b_2' \right) \biggr|_{x \to -\infty}^{x \to +\infty}
\nonumber
\\
&=& 8 i k_{\Omega} |R_{\Omega}|^2.
\end{eqnarray}
Eq. (\ref{Fermi-rule}) recovers the so-called {\rm Fermi golden rule} with 
$k_{\Omega} |R_{\Omega}|^2 > 0$ for all generic cases with nonvanishing coefficient $R_{\Omega}$.

Introducing the square amplitude $Q(\tau) = |c(\tau)|^2$, we obtain
a simple differential equation
\begin{equation}
\label{normal-form-growth}
K \frac{dQ}{d\tau} = - 8 \Omega k_{\Omega} |R_{\Omega}|^2 Q^2,
\end{equation}
starting with the positive initial value $Q(0)$. For the internal mode of
positive energy with $K > 0$, this equation leads to the slow 
(i.e., power law) decay
of the internal mode in time \cite{kiv2}. For the internal mode of
negative energy $K < 0$, this equation guarantees the power-law growth
of the internal mode in time and eventual blow up of the quadratic
approximation in (\ref{normal-form-growth}), although
this growth is typically saturated by the nonlinearity. Mathematical justification
of the normal form equation (\ref{normal-form-growth})
and the Fermi golden rule (\ref{Fermi-rule})
can by found in \cite{scipio,scipio1}.

{\it Case Example 1: anti-symmetric soliton in double-well potentials.}
The dynamics of the anti-symmetric (so-called $\pi$) solitons
in a double well potential has been explored
extensively in the recent physical literature (see
e.g.~\cite{malomedbook}) motivated by experiments in
atomic~\cite{markus,leblanc,zibold} and optical~\cite{zhigang,haelt}
physics. In comparison to the work presented herein,
such solitons were realized experimentally for short dynamical
time scales, for which no dynamical instability was detected \cite{zibold}. Here we report on
the weak (nonlinear) instability of the anti-symmetric solitons
in the NLS model (\ref{nls}) with the repulsive interaction $g = -1$
and the double-well potential
\begin{eqnarray}
V(x)=V_0 \left({\rm sech}^2(x-x_0) + {\rm sech}^2(x+x_0) \right),
\label{eqn4}
\end{eqnarray}
for which we take $V_0=-1$ and $x_0=2$. For $\omega=-0.4$, the frequency of
the internal mode of the anti-symmetric soliton
is $\Omega \approx 0.203$ (i.e., $\Omega < |\omega|$ but $2 \Omega > |\omega|$)
ensuring the second harmonic occurs inside the phonon band.

\begin{figure}[!ht]
\begin{center}
\includegraphics[width=0.45\textwidth]{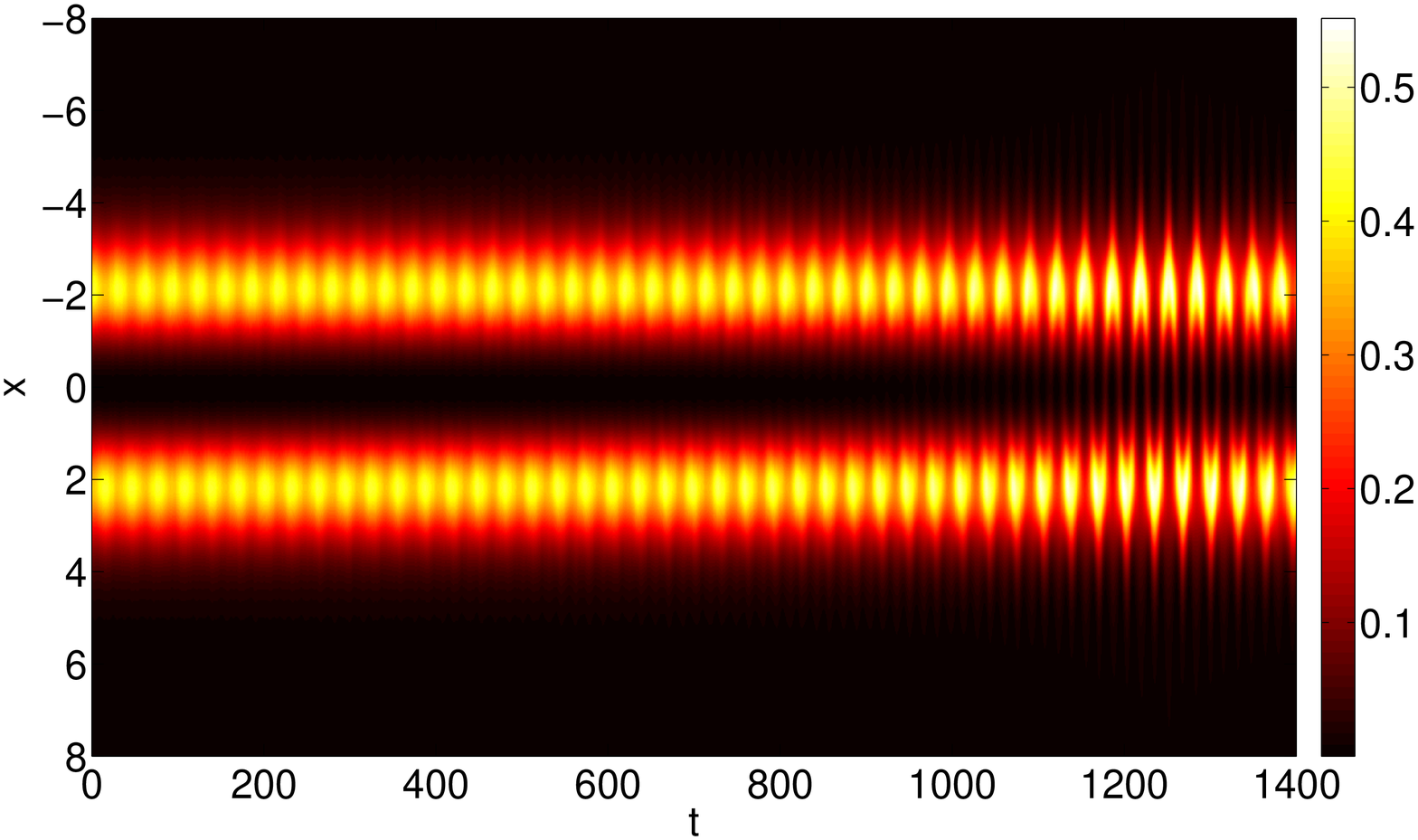}
\includegraphics[width=0.45\textwidth]{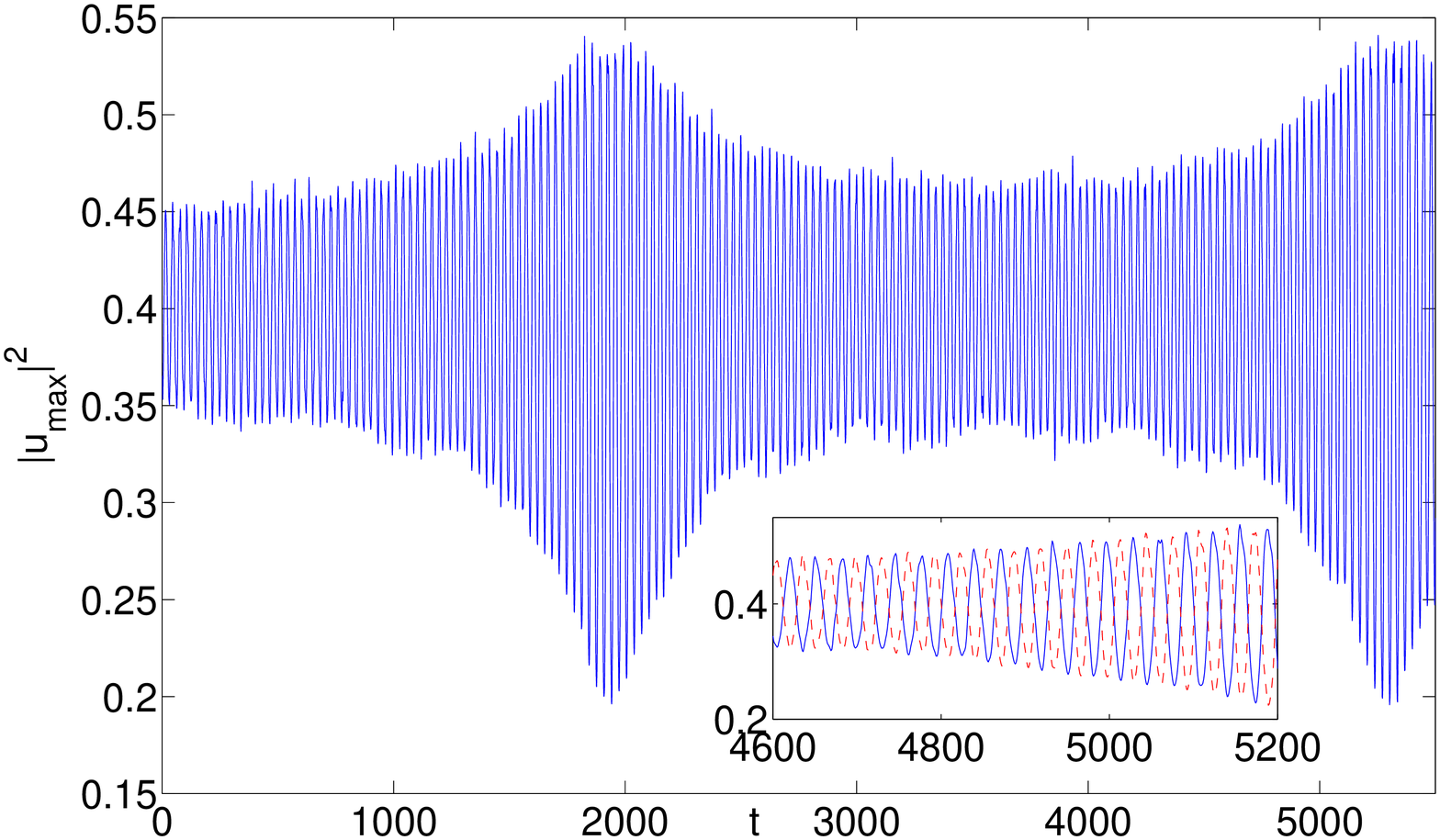}
\end{center}
\caption{(Color Online) The top panel shows the space-time evolution of a perturbed
anti-symmetric soliton in the NLS model (\ref{eqn1})
with the double-well potential (\ref{eqn4}). The slow growth over time
is further illustrated in the bottom panel of the figure containing
the evolution of the maximal amplitude in the left and right potential wells
by solid (blue) and dashed (red) line. Inset: detail of the growth
dynamics.}
\label{kps_fig2}
\end{figure}

In Fig.~\ref{kps_fig2}, we monitor the dynamical
evolution of the anti-symmetric soliton, perturbed by the internal mode.
We observe the slow manifestation
of a dynamical instability, both in the space-time evolution in the
top panel, as well as more concretely in the time evolution of the maximal
amplitudes in the two potential wells in the bottom panel.
The longer term dynamics shows that the growing oscillatory
dynamics of the internal mode does not saturate but
returns to the initial state leading to recurrent dynamics.

{\it Case Example 2: the twisted localized mode in a one-dimensional lattice with cubic nonlinearity.}
We examine a twisted localized mode in the one-dimensional cubic NLS
lattice~\cite{dnlsbook}. Such ``dipolar'' states have been previously
explored in both 1d~\cite{kip1d} and 2d~\cite{zhig2d} optical experiments.
The discrete NLS equation is a prototypical model of optical waveguide
arrays in nonlinear optics~\cite{motireview}. We take the discrete NLS equation in the
standard form:
\begin{eqnarray}
i \dot{u}_n = - C \Delta_2 u_n - |u_n|^2 u_n .
\label{eqn10}
\end{eqnarray}
Here $u_n$ plays the role of the envelope of the electric
field at the $n$-th waveguide and $C$ represents the
strength of the evanescent coupling between the waveguides,
while $\Delta_2$ stands for the standard 3-point stencil in
the one-dimensional lattice.

\begin{figure}[!ht]
\begin{center}
\includegraphics[width=0.45\textwidth]{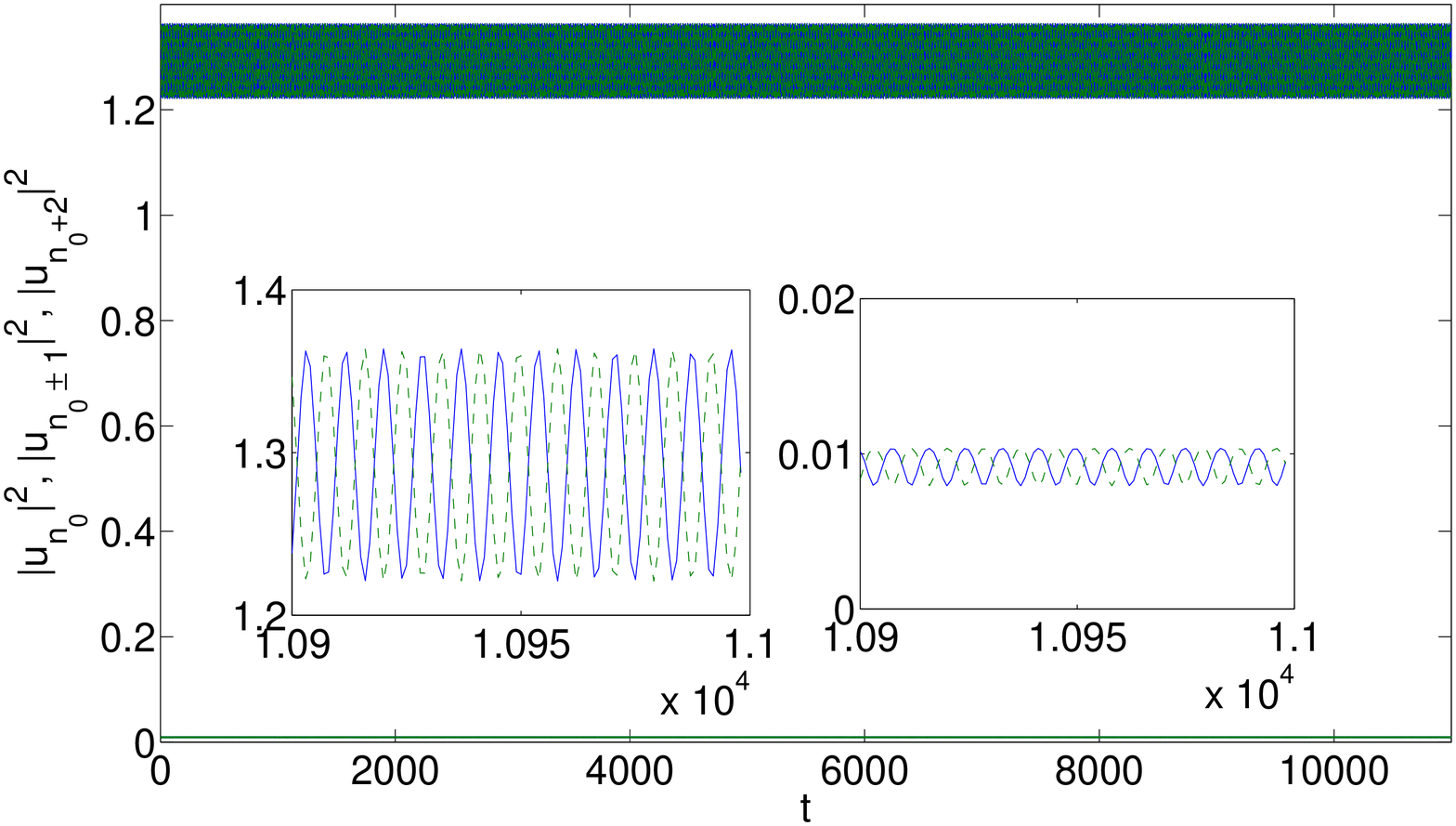}
\includegraphics[width=0.45\textwidth]{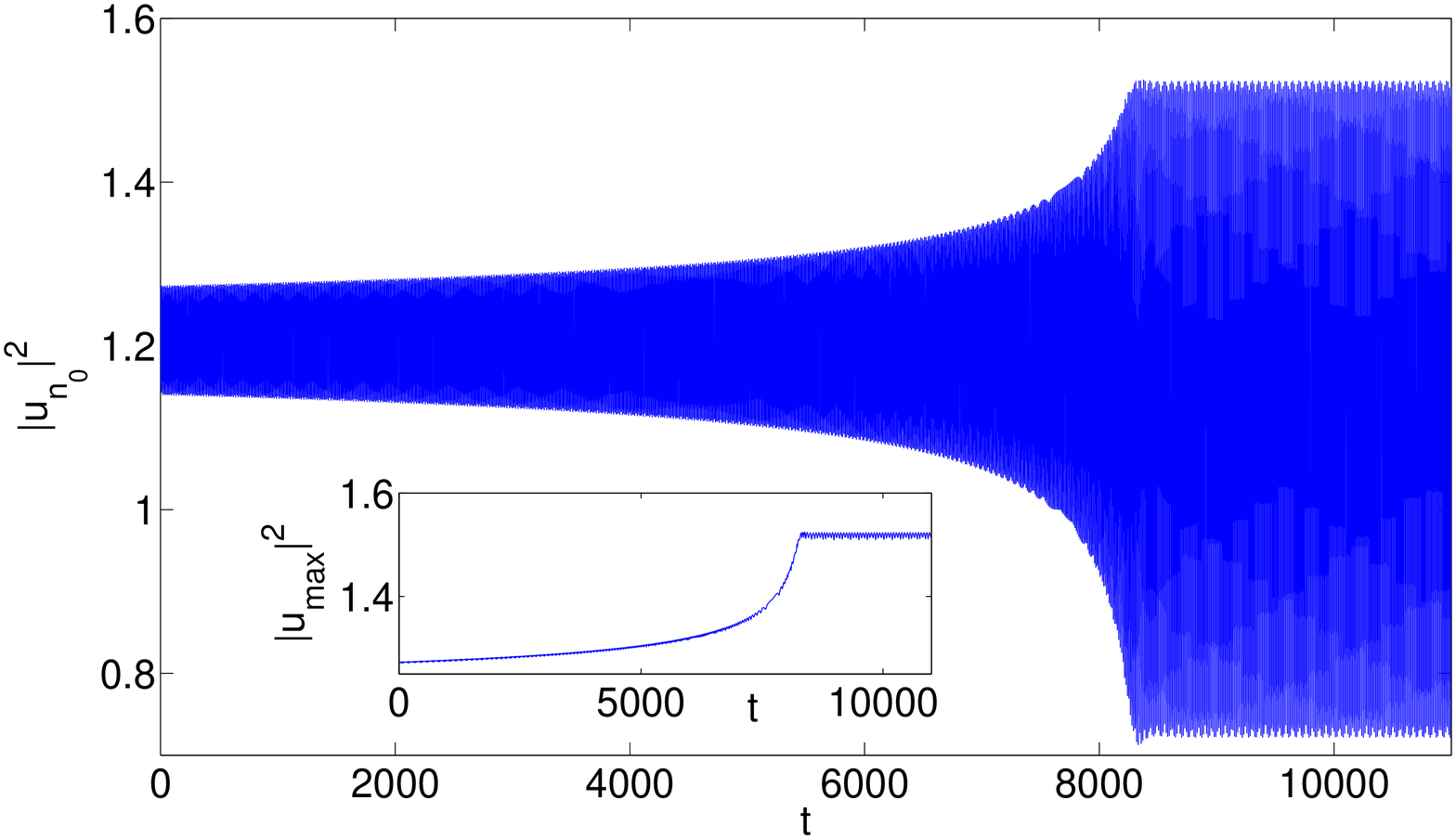}
\end{center}
\caption{(Color Online) Evolution of twisted localized modes in the discrete NLS model (\ref{eqn10}).
Top: time evolution of the 4 center-most sites (i.e., the two central
ones and their immediate neighbors) of the structure
for $C=0.01$; no instability is manifested. The insets show a zoom-in
of the relevant sites revealing their oscillatory behavior. Bottom:
the evolution of one of the central sites for $C=0.07$ (the inset shows
its envelope illustrating the unstable evolution). In both panels
$n_0=0$.
}
\label{kps_fig1}
\end{figure}

The twisted localized modes can be exactly represented in the
limit of $C=0$ (so-called anti-continuum limit)
as $u_n(t) = e^{i t} (\delta_{n,0}-\delta_{n,1})$, setting $\omega=-1$ without loss
of generality. As $C \neq 0$ increases, as shown in~\cite{pelikev1},
the solution is linearly stable but it has
one internal mode of negative energy with the frequency
$\Omega = \mathcal{O}(C^{1/2})$ as $C \to 0$. For small $C$, this frequency is
isolated from the phonon band, which corresponds to
the frequencies in the interval $[1,1+4C]$. For $C=0.01$, the internal mode frequency
is $\Omega \approx 0.204$ so that the condition  $2 \Omega > |\omega|$
is not satisfied.
While this case is unstable
too via the proposed mechanism, the instability only arises
through a higher harmonic (the {\it fifth harmonic}, in fact) of
the mode frequency. That is why the instability is not visible on the time
scale of our dynamical simulations shown in the top panel of Fig.~\ref{kps_fig1}.

On the other hand, for the case of $C=0.07$, the internal mode frequency
is $\Omega \approx 0.598$ and
$2 \Omega > |\omega|$
is satisfied. In the dynamics of the bottom panel of Fig.~\ref{kps_fig1}, 
we thus initialize
with such a solution, perturbed by this internal mode. We can
clearly see in this panel (showing the evolution of
one of the two central amplitudes at $n_0=0$) that there is very slow growth
reminiscent of a power law. After this instability manifests itself,
it eventually saturates. The theory
does not reveal any information about the
ultimate fate of the dynamics.
The numerics suggest a resulting genuinely
periodic state for the modulus, hence a genuinely quasi-periodic
(or breather-on-breather~\cite{kw}) state for the system.
This, in turn, suggests that it would be quite worthwhile
to further explore such states dynamically.

{\it Case Example 3: a discrete vortex in a two-dimensional saturable lattice.}
Our third example is also highly motivated
physically: we inspect discrete vortices in two-dimensional lattices that
were robust enough dynamically to also be accessible in
the optical observations within photorefractive optical crystals ~\cite{neshev2,moti2}.
Here, to comply with the
photorefractive nature of the nonlinearity and to illustrate
the genericity of our results a saturable nonlinearity
has been implemented~\cite{jiankenjp} (although the results
would still hold in the cubic case).
The discrete saturable NLS equation takes the form
\begin{eqnarray}
i \dot{u}_{n,m}=- C \Delta_2 u_{n,m} + \frac{u_{n,m}}{1+|u_{n,m}|^2},
\label{eqn5}
\end{eqnarray}
where $\Delta_2$ in this case stands for the standard 5-point stencil in the two-dimensional square lattice.

\begin{figure}[!ht]
\begin{center}
\includegraphics[width=0.45\textwidth]{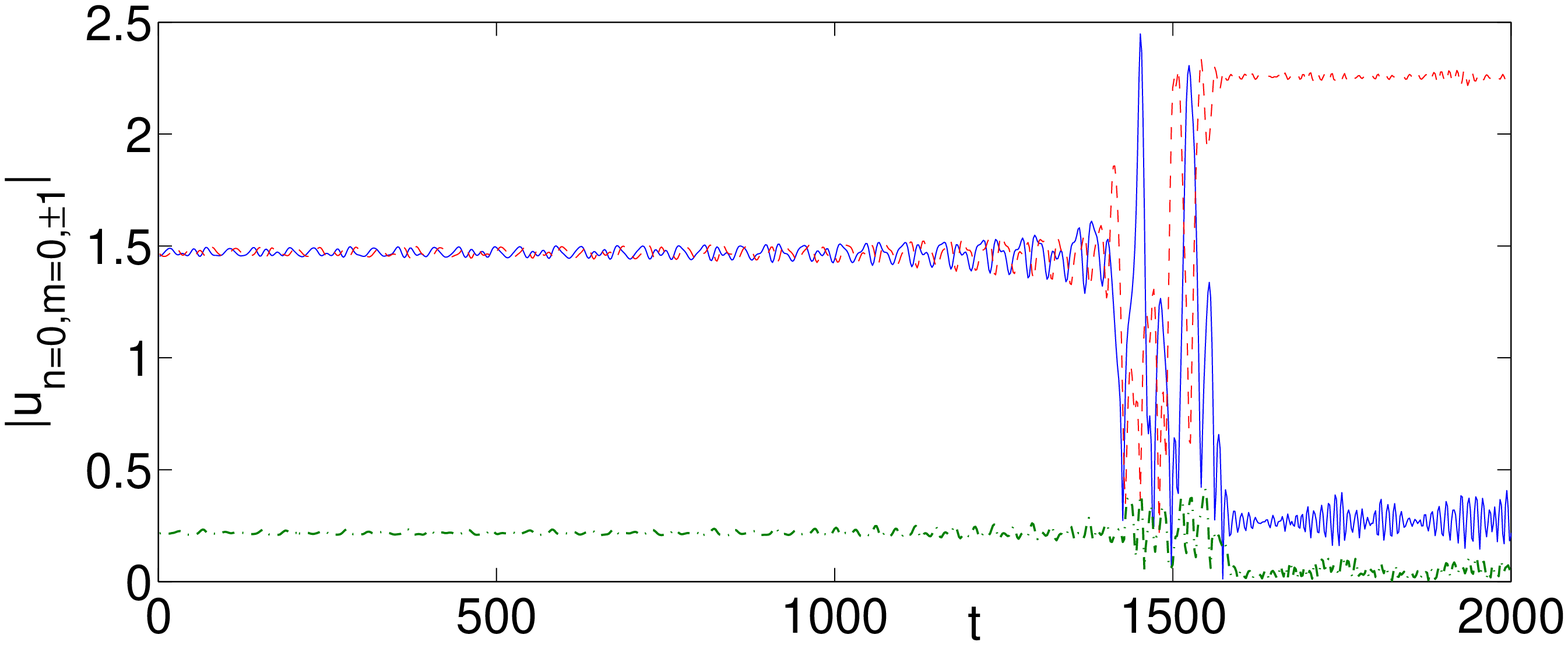}
\includegraphics[width=0.45\textwidth]{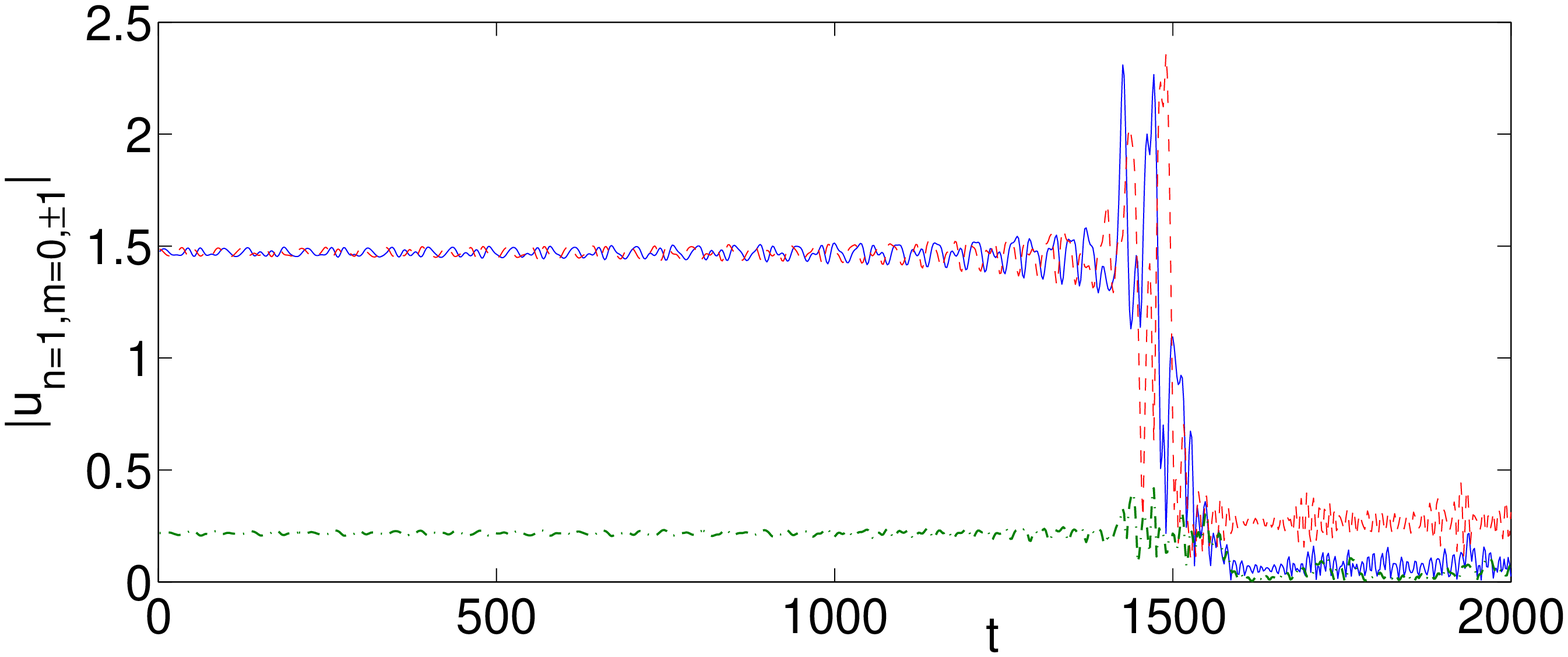}
\end{center}
\caption{(Color Online) The top panel
shows the modulus time evolution of $|u_{n=0,m=0,\pm 1}|$, while
the bottom panel illustrates the corresponding evolution of
$|u_{n=1,m=0,\pm 1}|$ [i.e., 4 sites constituting the discrete vortex
($(0,0)$, $(0,1)$, $(1,0)$ and $(1,1)$) 
and two immediately adjacent ones ($(0,-1)$ and $(1,-1)$)].
Among the 4 central sites, only one (dashed red in the top panel) grows
in power, while the other 3 (solid blue in the top and bottom and dashed red
in the bottom) decrease in amplitude. The neighboring sites also remain
at low amplitudes (green dash-dotted).}
\label{kps_fig3}
\end{figure}

Our example shown in Fig.~\ref{kps_fig3} corresponds to the case
of $C=0.09$. In this case, the phonon band corresponds to
the frequencies in the interval $[1-\omega,1-\omega+ 8 C]$. The
discrete vortex consisting of four excited principal sites
at the center of the chain with phases of approximately $0$,
$\pi/2$, $\pi$ and $3 \pi/2$ is found for $\omega=0.65$ to
possess three internal modes with eigenvalues $\Omega = 0.012$,
$\Omega = 0.167$ and $\Omega = 0.191$. Among the three,
and while the structure is linearly stable,
the last one has its second harmonic lying within the
phonon band, hence we anticipate its destabilization
by the mechanism reported herein.
Indeed, this is what we observe in Fig.~\ref{kps_fig3}.
However, the instability in this case appears to be far
more ``detrimental'' for the state in comparison to the
previous examples. In particular, the slow
growth of the instability  eventually gives rise to a dramatic event through
which one among the four vortex principal nodes picks up
most of the power in the system, while the other three
considerably decrease in power. The resulting 
dynamics effectively leads to a single-site ground state configuration
of the two-dimensional lattice.

{\it Conclusions.} In the present work, we examined a
previously unexplored mechanism for nonlinear instability which is
{\it generic} for excited states of nonlinear Schr{\"o}dinger
systems. 
The generality and
potential experimental ramifications (through numerical experiments herein)
of our findings were also discussed. It was shown that
the mechanism occurs independently of the discrete or continuum,
one- or multi-dimensional, cubic or saturable nature of the underlying NLS models,
as long as the excited nature of the state is manifested
via the internal modes of negative energy. These are
found to give rise to a weak (power law in its apparent manifestation)
instability that eventually deforms, or in some cases (e.g., the
discrete vortex) completely destroys the configuration.

While the mechanism presented herein is explicitly established,
there are numerous questions that are worthy of further
investigation, in addition to its potential experimental demonstration.
It would be relevant to confirm
that the mechanism can be numerically observed also in the case of higher
order harmonics e.g., when $n \Omega <  |\omega|$ but $(n+1) \Omega > |\omega|$.
Additionally, it would be relevant to explore
whether the mechanism is applicable to other dispersive wave
systems of intense current interest, such as, e.g., nonlinear
Dirac equations.


\begin{thebibliography}{99}



\bibitem{ablowitz1}
M.J. Ablowitz, B. Prinari and A.D. Trubatch,
{\it Discrete and Continuous Nonlinear Schr{\"o}dinger Systems},
Cambridge University Press (Cambridge, 2004).

\bibitem{sulem}
C. Sulem and P.L. Sulem,
\newblock {\it The Nonlinear Schr{\"o}dinger Equation},
Springer-Verlag (New York, 1999).

\bibitem{hasegawa}
A. Hasegawa, {\it Solitons in Optical
Communications}, Clarendon Press (Oxford, NY 1995).


\bibitem{kivshar}
Yu.S. Kivshar and G.P. Agrawal,
{\it Optical solitons: from fibers to photonic crystals},
Academic Press (San Diego, 2003).


\bibitem{zakh1}
V.E. Zakharov,
{Collapse and Self-focusing of Langmuir Waves},
\newblock Handbook of Plasma Physics, (M.N. Rosenbluth and R.Z. Sagdeev
eds.), vol. 2 (A.A. Galeev and R.N. Sudan eds.), 81--121, Elsevier (1984).

\bibitem{zakh2}
V.E. Zakharov,
{Collapse of Langmuir waves},
Sov.\ Phys.\ JETP {\bf 35} (1972) 908--914.

\bibitem{benjamin}
T.B. Benjamin and J.E. Feir,
J.\ Fluid Mech.\ {\bf 27}, 417 (1967).

\bibitem{onofrio}
M. Onorato, A.R. Osborne, M. Serio, and S. Bertone,
Phys.\ Rev.\ Lett.\ {\bf 86}, 5831 (2001).



\bibitem{becbook1}
L.P. Pitaevskii and S. Stringari,
{\it Bose-Einstein Condensation}, Oxford University Press (Oxford, 2003).



\bibitem{becbook2}
C.J. Pethick and H. Smith,
{\it Bose-Einstein condensation in dilute gases}, Cambridge University
Press (Cambridge, 2002).



\bibitem{rcg:BEC_BOOK}
P.G. Kevrekidis, D.J. Frant\-zes\-ka\-kis, and R. Carretero-Gonz{\'a}lez
{\sl Emergent Nonlinear Phenomena in Bose-Einstein Condensates: Theory and Experiment}.
Springer Series on Atomic, Optical, and Plasma Physics, Vol.~{\bf 45}
(Heidelberg, 2008).




\bibitem{toddkeith} T. Kapitula, and K. Promislow,
{\it Spectral and dynamical stability of nonlinear waves},
Springer-Verlag (New York, 2013).

\bibitem{jyang} J. Yang,
{\it Nonlinear Waves in Integrable and Nonintegrable Systems}.
SIAM (Philadelphia, 2010).

\bibitem{pelinov} D.E. Pelinovsky,
{\it Localization in Periodic Potentials},
Cambridge University Press (Cambridge, 2011).

\bibitem{dnlsbook} P.G. Kevrekidis,
{\it The Discrete Nonlinear Schrödinger Equation}
Springer-Verlag (Heidelberg, 2009).

\bibitem{kapkev} T. Kapitula, P.G. Kevrekidis, and B. Sandstede, Phys. D {\bf 195} 263 (2004).



\bibitem{kiv1} Yu.S. Kivshar, D.E. Pelinovsky, T. Cretegny, and M. Peyrard, 
Phys. Rev. Lett. {\bf 80}, 5032--5035 (1998)

\bibitem{kiv2} D.E. Pelinovsky, Yu.S. Kivshar, and V.V. Afanasjev, 
Physica D {\bf 116}, 121--142 (1998).

\bibitem{scipio} S. Cuccagna, Phys. D {\bf 238}, 38 (2009).

\bibitem{scipio1} S. Cuccagna, and M. Maeda, arXiv:1309.0655.



\bibitem{malomedbook} B.A. Malomed (Ed.),
{\it Spontaneous symmetry-breaking, self-trapping and Josephson
oscillations}, Springer-Verlag (Heidelberg, 2013).



\bibitem{markus} M. Albiez, R. Gati, J. F\"{o}lling, S. Hunsmann, M.
Cristiani, and M. K. Oberthaler,
Phys. Rev. Lett. \textbf{95}, 010402 (2005).


\bibitem{leblanc} L. J. LeBlanc, A. B. Bardon, J. McKeever, M. H. T. Extavour, D. Jervis, J. H. Thywissen, F. Piazza, and A. Smerzi,
Phys. Rev. Lett. {\bf 106}, 025302 (2011).


\bibitem{zibold} T. Zibold, E. Nicklas, C. Gross,
and M.K. Oberthaler,
Phys. Rev. Lett. {\bf 105}, 204101 (2010).

\bibitem{zhigang} P.G. Kevrekidis, Z. Chen, B. A. Malomed, D. J.
Frantzeskakis, and M. I. Weinstein,
 Phys. Lett. A \textbf{340}, 275 (2005).

\bibitem{haelt} C. Cambournac, T. Sylvestre, H. Maillotte , B.
Vanderlinden, P. Kockaert, Ph. Emplit, and M. Haelterman,
Phys. Rev. Lett.
\textbf{89}, 083901 (2002).

\bibitem{kip1d} M. Stepi{\'c}, E. Smirnov, C. R{\"u}ter,
L. Pr{\"o}nneke, D. Kip, and V. Shandarov, Phys. Rev. E {\bf 74}, 046614 (2006).

\bibitem{zhig2d} J. Yang, A. Bezryadina, I. Makasyuk, and Z. Chen,
Opt. Lett. {\bf 29}, 1662 (2004).


\bibitem{motireview} F. Lederer, G.I. Stegeman, D.N. Christodoulides,
G. Assanto, M. Segev, and Y. Silberberg,
Phys. Rev. {\bf 463}, 1 (2008).

\bibitem{pelikev1} D.E. Pelinovsky, P.G. Kevrekidis, and D.J. Frantzeskakis,
Phys. D {\bf 212}, 1 (2005).

\bibitem{kw} P.G. Kevrekidis, and M.I. Weinstein, Math. Comp. Simul. {\bf 62}, 65 (2003).

\bibitem{neshev2} D.N.Neshev, T.J. Alexander, E.A. Ostrovskaya,
Yu.S. Kivshar, H. Martin, I. Makasyuk, and Z. Chen,
Phys. Rev. Lett. {\bf 92}, 123903 (2004).


\bibitem{moti2} J.W. Fleischer, G. Bartal, O. Cohen, O. Manela,
M. Segev, J. Hudock and D.N. Christodoulides,
Phys. Rev. Lett. {\bf 92}, 123904 (2004).


\bibitem{jiankenjp} J. Yang,
New J. Phys. {\bf 6}, 47 (2004).



\end{thebibliography}
\end{document}